# Extreme Elastic Deformation of Atoms and Pressure-Induced Superconductivity in Silicon

Xiaozhi Hu

Department of Mechanical Engineering

University of Western Australia

Perth, WA 6009, Australia

xiao.zhi.hu@uwa.edu.au

**Abstract**

Change in the interatomic spacing of a two-atom system under tension and compression has been modelled by the elastic deformation of atoms. The critical elastic strain of atoms before separation or cracking from tension was estimated by the Griffith theory together with a recent mechanics model, then extended to the lateral elastic expansion under uniaxial compression. The hypothesis of deformable atoms has led to astonishing predictions of the critical elastic strain, around 10 and 20% for silicon in the {110} and {100} crystal directions. Superimposed by the substantial reduction of interatomic spacing in the direction of uniaxial compression above 20 GPa, these severely deformed silicon atoms or metastable "new variants" have acquired unforeseeable characteristics and properties, vastly different from those of silicon atoms under moderate stresses. Under extreme pressure, the natural repulsive reaction of atoms is intensified due to the increasing alignment of electron orbitals along the pressure direction and formation of metastable pressure-resistant phases. An opportunity for creation of a superconductive band has thus arisen at the edge of the laterally elastically expanded region away from the nuclei, where more space is available for free electron movement. Diamond results were also used to validate the new mechanics model, including the effects of atomic scale defects on fracture strain and strength, critical to elastic strain engineering.

## 1. Introduction

If only one sentence could be passed to the next generations of scientists, "I believe it is the atomic hypothesis that all things are made of atoms – little particles that move around in perpetual motion, attracting each other when they are a little distance apart, but repelling upon being squeezed into one another" [Feynman 1963], as quoted in Muser et al. [2022]. Why do atoms naturally react to external forces in the way Feynman described if electrons are randomly orbiting around the nuclei? Different reactions of atoms under tension and compression indicate the random electron orbitals pertinent to the outer shell of atoms must have reacted differently to different stresses. This is only possible if atoms have been deformed under stress, then electrons inside those deformed atoms have to react to the changing space and stress environments. The properties and formation of nuclei cannot be changed by simply applying pressure, so the pressure-induced superconductivity in silicon and other hard solids most likely will be linked to some rearrangements of electron orbitals due to the elastic deformation of atoms under severe pressure.

It is certain that repulsion of atoms must be present when the pressure-induced superconductivity in silicon and other brittle solids is observed [Lin et al. 1986, McMillan 2002, Ekimov et al. 2004, Lorenz and Chu 2005, Kim et al. 2018, Somayazulu et al. 2019, Lim et al. 2022]. This extraordinary phenomenon is typically witnessed under extremely high pressures from around 20 GPa to close to 200 GPa. The expected reduction in the inter-atomic spacing under extreme pressure as we learn from the Feynman's description implies atoms probably have been deformed. The magnitude of atomic deformation and the subsequent inevitable influence on electron orbitals inside the atoms could hold the key to revealing the mechanism of pressure-induced superconductivity. The dramatic change in the space around the nuclei of severely deformed metastable atoms will generate unimaginable impacts on electron orbitals and electron movements. To explore the unknowns of "deformed atoms" under extreme pressure, the hypothesis that atoms can be elastically deformed before separation or fracture is investigated and modelled around the critical elastic strain $\varepsilon_c$. Hopefully, the current study can provide fresh interpretations on the pressure-induced superconductivity from the viewpoint of elastic deformation of atoms.

"Elastic strain engineering for unprecedented materials properties" by Li et al. [2014] outlines unique conditions of extreme elastic strains and exciting opportunities for unprecedented findings within the elastic limit. As such, the main objective of "Elastic Strain Engineering" (ESE) is to explore unusual physical and chemical properties, near but below the critical elastic strain $\varepsilon_c$, that are radically different from the common properties of materials experiencing only moderate stresses. It has been postulated [Li et al. 2014] that the critical elastic strain $\varepsilon_c$ is linked to a characteristic length of a material.

$$\varepsilon_c \propto \frac{1}{L^\alpha} \qquad 0.5 \leq \alpha \leq 1 \tag{1}$$

Here $L$ can be considered as the dominant characteristic length scale of a material consisting of various levels of hierarchical microstructures. The exponent $\alpha$ is between 0.5 and 1.0 [Li et al. 2014, Zhu and Li 2010]. At the atomic scale, the characteristic length $L$ most likely is linked to the atomic diameter for single crystals.

Eq. (1) on the critical elastic strain $\varepsilon_c$ is applicable to a wide range of materials. Taking a nanocrystalline material as an example, $L$ is the average grain size rather than the size of crystal unit cells. Such a definition of the dominant microstructure has been adopted for metals many years ago by the Hall-Petch relation [Hall 1951, Petch 1953, Yip 1998, Armstrong 2014]. For layered nano-/micro- structures, $L$ can be the thin film thickness [Li et al. 2014] or a combined measurement of different nano-layers in superlattices [Chen et al. 2023], or even the micro-layer thickness in micro-/macro- laminar composites [Hu et al. 2021].

Theoretically, the critical elastic strain $\varepsilon_c$ of single crystal silicon (SCS), before cracking, will be the upper limit of lateral elastic expansions generated under extreme uniaxial compression. In order to avoid cracking, the critical elastic strain $\varepsilon_c$ also becomes a practical concern in experiments for the pressure-induced superconductivity. Obviously, the elastic deformation of a bulk SCS component under ambient temperatures and pressures is closely linked to the elastic properties of crystal unit cells with known packing orders of silicon atoms, as commonly described by the "ball and stick" model [Pasquarello et al. 1998, DelRio et al. 2015]. While it is well known that elastic properties of crystalline solids depend on the size and shape of unit cells and lattice parameters, it seems the critical elastic strain and deformation of individual atoms under either tension or compression have not been explicitly calculated. It has been observed that high-pressure phases of silicon [Clarke et al. 1988, Mizushima et al. 1994, Bradby et al. 2001, Zhang and Zarudi 2001] are formed under extreme compression. While the simple "ball and stick" model commonly adopted for molecular dynamics simulations and illustrations of crystal structures [Pasquarello et al. 1998, DelRio et al. 2015] may be able to model sliding of atoms under compression, phase transformation processes under high pressures and possible elastic deformation of atoms cannot be clearly explained. It is highly likely the phase transformation process under high pressures will involve changes in the boundaries of atoms (in both lateral and compressive load directions) and alterations of previously random electron orbitals within those "new metastable atoms", that are drastically different to the original silicon atoms at ambient pressures. It will be curious to see whether the simple and important relation specified in Eq. (1) is still valid at the atomic scale for SCS under extreme tensile or compressive stresses. Clearly, Elastic Strain Engineering (ESE) of brittle solids and compounds for special pressure-induced physical and chemical effects cannot be fully understood and successfully executed without a good understanding of the extreme elastic properties at the atomic scale. The effect of "smaller is stronger" [Li et al. 2014, Zhu and Li 2010] also needs to be tested at the lowest size limit – the atomic scale.

Bearing in mind the potential application of Eq. (1) at the atomic scale, the objective of this study is to testify the hypothesis that atoms can be elastically deformed under external stresses and to tentatively define the critical elastic deformation of silicon atoms under uniaxial tensile stresses using a simple mechanics model, then link it to the corresponding critical elastic strain in the x-y plane ($\sigma_x \approx \sigma_y \approx 0$) under extreme z-directional pressure. Through identifying a plausible elastic-strain based mechanism for superconductivity, hopefully this simple mechanics model and the estimated critical elastic strain and relevant deformation of atoms may shed light on the pressure-induced superconductivity in silicon and similar solids, or on other extraordinary physical properties pertinent to the critical elastic strain $\varepsilon_c$.

## 2. Extreme Elastic Deformation of Atoms and Superconductivity

### 2.1 Critical elastic strain $\varepsilon_c$ of silicon atoms from uniaxial tension

Elastic properties of single crystal silicon (SCS) are influenced by the crystal unit cell structures and crystallographic orientations. For a given crystal direction, the bulk elastic deformation of a SCS component has to come from cumulative effects from all unit cells involved. Zooming in on one unit cell, its elastic properties and deformation are fully determined by the silicon atoms within the unit cell and their covalent bonds at the mutual boundaries. Most likely, the elastic deformation of a unit cell would come mostly from the elastic deformation of atoms if a space-filled atomic model were assumed [Wikipedia, Space-filling model]. Although it may be hard to envisage silicon atoms (only around 0.235 nm) can be elastically deformed under stress, a deformable atom model is still far realistic than closely-packed "rigid" atoms only capable of sliding and rotating. In fact, the commonly used "ball and stick" model [Pasquarello et al. 1998, DelRio et al. 2015] suggests atoms are deformable if the "sticks" (representing the covalent bonds) are changed under stress either in length or angle. If the space-filled atomic model is used, any change in the interatomic spacing means the atoms have been deformed because the space where deformation has occurred is still filled, and the atomic boundaries are still intact either under tension or compression.

At the atomic scale, SCS has an orderly packing pattern [Li et al. 2003], as shown in Fig. 1(a), which is ideal for estimation of the critical elastic deformation of atoms. While the Bohr atom model with fixed electron orbitals and energies as illustrated in Fig. 1(a) suggests a well-defined atomic radius under 0-stress, the Schrodinger quantum model with random electron orbitals around the nucleus can only indicate a statistically averaged atomic radius [McKagan et al. 2008, Bohr 1923, Schrodinger 1926]. In this study, the nearest-neighbour distance between two atoms (two nuclei) [Sinclair and Lawn 1972, Field 2012] is taken as the average atomic diameter $D_a$ of silicon in the equilibrium state without pressure or tensile stress. It can be imagined that in the quantum world of atoms and electrons, nuclei resonate with randomly orbiting electrons, so atoms are continuously oscillating around their equilibrium positions in the form of elastic deformation. Since atoms are bonded together through electron sharing, the constant oscillation means the shape and size of all atoms and their boundaries are constantly changing, and so is the inter-atomic spacing. In other words, if the Schrodinger quantum model is adopted in modelling, it is already conceded indirectly that all atoms can be deformed elastically. The stress-induced elastic deformation of atoms then comes naturally. In fact, the atomic vibration illustrated in Fig. 1(b) has been recognised for many years [Dean 1966], except it is not explicitly linked to the elastic deformation of atoms. So, the critical elastic strain $\varepsilon_c$ of single crystals and extreme elastic deformation of atoms have not been linked together. Nowadays, the critical elastic strain $\varepsilon_c$ of hard solids is emphasized for Elastic Strain Engineering (ESE) in order to discover extraordinary physical and chemical properties associated with extreme elastic deformation.

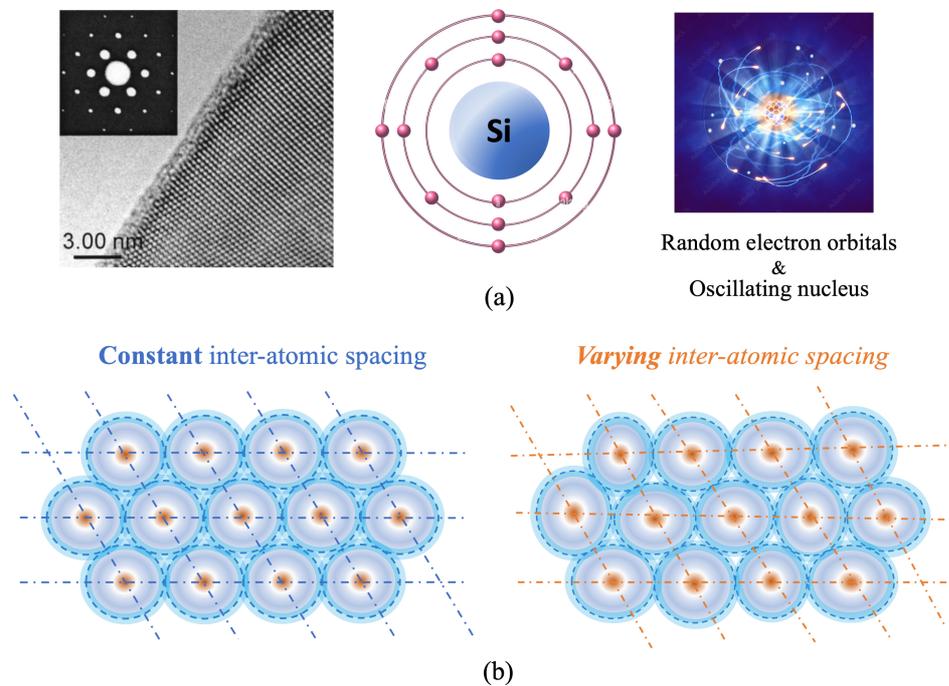

Figure 1. (a) Orderly packed silicon atoms in SCS [Li et al. 2003]. Bohr atom model defines constant inter-atomic spacing (under 0-stress). Schrodinger quantum model with random electron orbitals implies the inter-atomic spacing is constantly changing (even under 0-stress). (b) In the quantum world of atoms and electrons, nuclei resonate with randomly orbiting electrons, so atoms are continuously oscillating around their equilibrium positions in the form of elastic deformation. Stress-induced elastic deformation of atom comes naturally for the Schrodinger quantum model.

The hypothesis of deformable atoms is illustrated in Fig. 2. Before applying any stress on the two-atom system, the equilibrium distance between the two atoms is the atomic diameter $D_a$ or the nearest-neighbour distance between the two nuclei. The atomic boundary with the covalent bond is formed through shared electrons, and is about the midpoint between the two nuclei. Close to the critical elastic limit with the applied stress $\sigma \to \sigma_{th}$, the two atoms under tension are still bonded together but with the enlarged critical diameter $D_{a\text{-}c}$ because of the increase in inter-atomic distance or due to the elastic deformation of atoms. Under compression, the elastic deformation is limited by the lower limit $D_0$ (determined by the nuclei properties and formation of pressure-resistant metastable phases) as the atomic repulsion force is getting stronger and stronger close to the nuclei. In a way, the deformable atom model illustrated in Fig. 2 may represent a combination of the common "ball and stick" and "space-filled" atomic models, but with a keen focus on the well-defined atomic boundary formed by the strong covalent bond from shared electrons and the elastic deformation of atoms generated by external stresses. Clearly, a space-filled atomic structure is only elastically deformed under moderate stresses, so the boundaries of atoms remain intact.

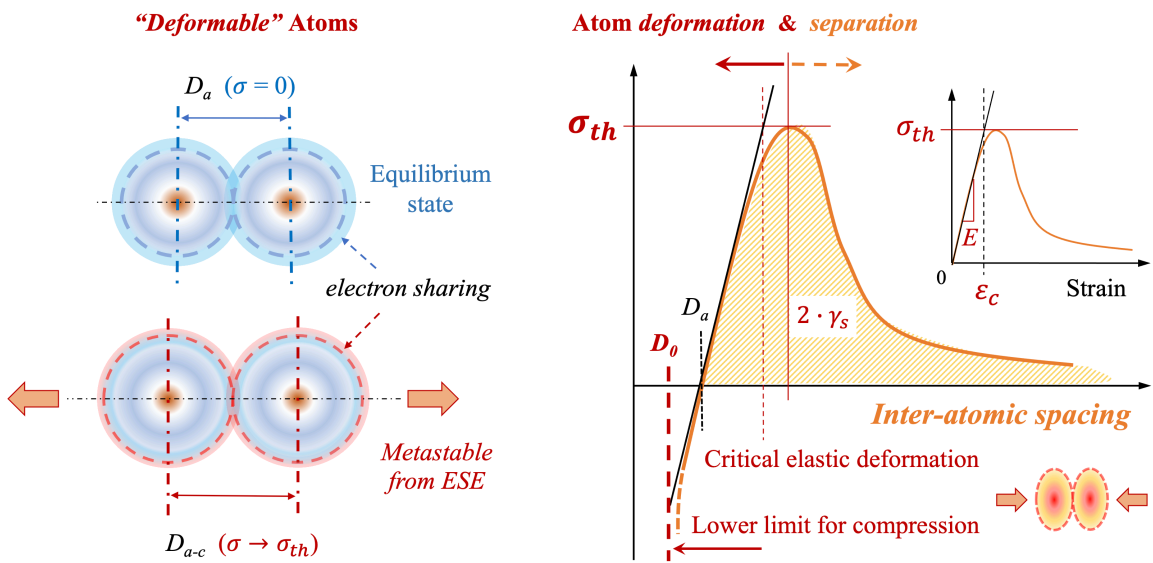

Figure 2. Inter-atomic spacing (under 0 stress) = atomic diameter $D_a$ (averaged equilibrium state for the nearest-neighbour distance). Any change in the inter-atomic spacing indicates atoms have been elastically deformed. $D_{a\text{-}c}$ is the average diameter of critically elastically deformed atoms from tension. The theoretical strength $\sigma_{th}$ is the criterion for separation of atomic bonds, linked to the elastic modulus $E$ and critical elastic strain $\varepsilon_c$ of atoms by the mathematical definition $\sigma_{th} = E \cdot \varepsilon_c$. The lower limit $D_0$, determined by properties of the nucleus and formation of new metastable pressure-resistant phases, limits the compressive deformation. Natural repulsive reaction of atoms against further compression towards $D_0$ leads to the non-linear compressive behaviour.

Separation or fracture of silicon atoms occurs when the interatomic spacing passes the critical elastic limit corresponding to the theoretical strength $\sigma_{th}$ and the critical strain $\varepsilon_c$, as illustrated in Fig. 2. The Griffith theory [1921] for brittle fracture links the fracture toughness $K_{IC}$ of a solid to its surface energy $\gamma_s$, Poisson's ratio $\nu$ and elastic modulus $E$, as in Tanaka et al. [2003].

$$K_{IC} = \sqrt{\frac{2 \cdot E \cdot \gamma_s}{1-\nu^2}} \tag{2}$$

It has been shown recently [Hu et al. 2022, Hu 2022] the fracture toughness $K_{IC}$ of single crystal silicon (SCS) can also be linked to its atomic diameter $D_a$ and theoretical strength $\sigma_{th}$ for a given crystal orientation.

$$K_{IC} = 2 \cdot \sigma_{th} \cdot \sqrt{3 \cdot D_a} \tag{3}$$

The nearest-neighbour distance between two atoms or between two nuclei to be more precise (around 0.235 nm) is taken as the atomic diameter $D_a$ of silicon [Sinclair and Lawn 1972, Field 2012] in the equilibrium state without pressure or tensile stress.

Independently, neither Eq. (2) nor Eq. (3) can do much besides their usual application for $K_{IC}$ estimations, but a surprise is just around the corner. The following relation can be established from Eq. (2) and (3) by introducing a correction factor $C_K$ (around one) since theoretical predictions of the fracture toughness $K_{IC}$ from different models may vary to some extent.

$$C_K \cdot \sqrt{\frac{2 \cdot E \cdot \gamma_s}{1-\nu^2}} = 2 \cdot \sigma_{th} \cdot \sqrt{3 \cdot D_a} \tag{4a}$$

For simplicity, for the time being it can be assumed that $C_K = 1$ so that the above relation can be rearranged into the following form:

$$\frac{\gamma_s}{1-\nu^2} = 2 \cdot 3 \cdot \left(\frac{\sigma_{th}}{E} \cdot D_a\right) \cdot \sigma_{th} \tag{4b}$$

Mathematically, $\sigma_{th}/E$ is the critical elastic strain $\varepsilon_c$ of a silicon atom based on the elastic stress and strain definition, as illustrated in the insert of Fig. 2. Although an atom may not be uniformly stretched across the electron clouds of different energy bands, the averaged critical elastic strain $\varepsilon_c$ of extremely deformed atoms (just before separation or fracture) can still be approximated as follows:

$$\varepsilon_c = \frac{\sigma_{th}}{E} = \frac{1}{6 \cdot (1-\nu^2)} \cdot \frac{\gamma_s}{\sigma_{th}} \cdot \frac{1}{D_a} \tag{5a}$$

The atomic diameter $D_a$ in the above equation can be taken as the dominant characteristic length scale of SCS, or $D_a = L$ in Eq. (1). Then Eq. (5a) shows the critical elastic strain $\varepsilon_c \propto D_a^{-1}$ with the exponent $\alpha = 1$. It is astonishing to see Eq. (1) is indeed valid at the atomic scale and has such a simple and well-defined relation.

While the theoretical strength $\sigma_{th}$ is linked directly to the critical elastic strain $\varepsilon_c$, the surface energy $\gamma_s$, as can be seen from Fig. 2, covers the entire energy dissipation process from the initial elastic deformation of atoms before $\sigma_{th}$ to the final separation after $\sigma_{th}$. Yet both are important to the critical elastic strain $\varepsilon_c$ as shown in Eq. (5a). It should also be mentioned that before the atomic attraction reaches the theoretical strength $\sigma_{th}$, the two atoms are still bonded together having the common boundary with shared electrons.

The critical elastic deformation of an atom before separation or cracking, $D_{a-c} - D_a$, as illustrated in Fig. 2 can then be determined from Eq. (4b) or Eq. (5a), i.e.,

$$\Delta D_a = D_{a-c} - D_a = \frac{\sigma_{th}}{E} \cdot D_a = \frac{1}{6 \cdot (1 - v^2)} \cdot \frac{\gamma_s}{\sigma_{th}} \tag{5b}$$

It is somewhat unexpected that the critical elastic deformation of an atom is independent of the atomic diameter. This would be unthinkable for the bulk elastic deformation of a silicon component as the total elastic deformation has to be linked to the component size and its engineering strain. One plausible explanation could be the critical atomic deformation before separation is around a common electron energy band width, which may be the extreme limit for atomic bonding. While the predictions from Eq. (5a) and (5b) are still to be verified, they will certainly stimulate further thinking. If proven correct, these theoretical predictions will promote more focused experiments and observations into those severely elastically deformed atoms with amended electron orbitals.

Table 1. Properties of single crystal silicon with atomic diameter $D_a = 0.235$ (nm)

|   | $\sigma_{th}$ (GPa) | $\gamma_s$ (J·m$^{-2}$) | $E$ (GPa) | Poisson's ratio $v$ |
|---|---|---|---|---|
| {100} | 26.4 | 2.55 | 130 | 0.279 |
| {110} | 17.0 | 1.80 | 169 | 0.362 |
| {111} | 21.0 | | | |

The major material properties of SCS in {100} and {110} planes are listed in Table 1. The theoretical strength $\sigma_{th}$ of SCS is from Dubois et al. [2006], the surface energy $\gamma_s$, the elastic modulus $E$ and Poisson's ratio $v$ are from DelRio et al. [2015]. The atomic diameter $D_a = 0.235$ (nm) is assumed to be the nearest neighbour distance [Sinclair and Lawn 1972, Field 2012]. The predicted fracture toughness $K_{IC}$ from Eq. (1) and Eq. (2) are listed in Table 2 together with the indentation fracture toughness measurements from Tanaka et al. [2003]. As already known (e.g. Tanaka et al., 2003), Griffith theory tends to predict the lower limit of $K_{IC}$, which will affect the predictions from Eq. (4b) and (5) because $C_K > 1$. For comparison purposes, all the critical elastic strain values from $\sigma_{th}$ and $E$ calculations and estimations using the surface energy $\gamma_s$ and the theoretical strength $\sigma_{th}$ and the atomic diameter $D_a$ are listed in Table 3.

The results in Table 3 show that the critical elastic strains predicted by Eq. (5a) using $\gamma_s$, $\sigma_{th}$ and $D_a$ are smaller than the estimations from $\sigma_{th}$ and $E$. The sign > is used in Table 3 since lower $K_{IC}$ is predicted from Eq. (2). The two critical elastic strain $\varepsilon_c$ predictions for the {110} plane in Table 2 are close to each other because of the reasonably similar $K_{IC}$ values. If the indentation fracture toughness around 1.29 (MPa√m) is used to correct the Griffith prediction, $C_K = 1.29/0.848$ should be used in Eq. (4a). As a result, the critical elastic strain $\varepsilon_c$ predicted from $\gamma_s$ and $\sigma_{th}$ and $D_a$ using Eq. (5a) is larger than 17.2% for the {100} plane. The critical elastic strain $\varepsilon_c$ of silicon atoms around 10% or 20% is extraordinarily large in comparison with the elastic strain of bulk silicon, which is typically < 1%.

Table 2. Fracture toughness $K_{IC}$ from three different methods

| **Methods** | **Eq. (2)** | **Eq. (3)** | Indentation |
|---|---|---|---|
| {100} $K_{IC}$ (MPa√m) | 0.848 | 1.40 | 1.29 |
| {110} $K_{IC}$ (MPa√m) | 0.837 | 0.903 | 1.12 |
| {111} $K_{IC}$ (MPa√m) | | 1.12 | |

Table 3. Critical elastic strain $\varepsilon_c$ estimated from the two definitions in Eq. (5a)

| | $\sigma_{th}/E$ | **Critical elastic strain $\varepsilon_c$** from $\gamma_s$ and $\sigma_{th}$ and $D_a$ | **Critical elastic deformation $\Delta D_a$ (nm)** |
|---|---|---|---|
| {100} | 20.3% | > 7.43% | > 0.0175 |
| {110} | 10.1% | > 8.64% | > 0.0203 |

The extremely large critical elastic deformation around 10 – 20% can impose substantial changes to the electron orbitals and electron movements inside those severely deformed atoms. Because of the extreme elastic strain and deformation, the metastable silicon atoms under severe stresses probably can be considered as "new silicon atoms" with drastically different properties compared to those original stress-free atoms. It is foreseeable that fewer electrons would be at the boundary region of two atoms with the critical atomic diameter $D_{a-c}$ and there would be even fewer electrons around the boundary if the temperature is dropped down to only few degrees *K*. These conditions are obviously favourable to the stress-induced superconductivity phenomenon. As a first approximation, it can be assumed that the three energy levels in Fig. 1(a) had the equal spacing, the electron energy band width would be 0.039 nm (= 0.235/6). The 10-20% critical elastic strain can create a new band width between 0.0235 – 0.047 nm, which is about one average energy band width. If confirmed, this will definitely have a huge impact on the physical properties of silicon atoms.

It is quite overwhelming to see the current simple analysis indicates the critical elastic strain $\varepsilon_c$ of silicon atoms can be as high as 10% or 20% in the {110} and {100} crystal planes. However, there are other reports in the literature, showing large elastic strains from numerical simulations. For instance, the early modelling work on engineering strains from Dubois et al. [2006] also suggested similar or even higher strains in {110} and {100} directions, possibly because the non-linear elastic section up to the theoretical strength $\sigma_{th}$ (illustrated in Fig. 2) was also included. Tensile tests of silicon nanowires, around 100 (nm) in diameter, performed by Zhang et al. [2016] proved the elastic strain at the fracture point indeed could be as high as 10 – 16% along the {110} crystal plane. They noticed the presence of the nano-scale surface roughness and processing defects limited the elastic deformation and fracture strength of those SCS nanowires. This problem will be dealt with in detail in Section 4.3.

2.2  Superconductivity created from lateral elastic expansion

Consider a silicon thin plate sample under the tri-axial stresses, $\sigma_x$ and $\sigma_y$ and $\sigma_z$. At the critical elastic limit, the tri-axial stresses can be linked to its theoretical strength $\sigma_{th}$, i.e.,

$$(\sigma_x - \sigma_y)^2 + (\sigma_y - \sigma_z)^2 + (\sigma_z - \sigma_x)^2 = 2 \cdot \sigma_{th}^2 \tag{6}$$

For the uniaxial compressive condition adopted for pressure-induced superconductivity experiments, it can be assumed that $\sigma_x = \sigma_y = 0$. The elastic strain in the x and y direction is given by:

$$\varepsilon_x = \varepsilon_y = \frac{1}{E} \cdot [\sigma_x - \nu \cdot (\sigma_y + \sigma_z)] = \frac{\sigma_{th}}{E} \cdot \left(-\nu \cdot \frac{\sigma_z}{\sigma_{th}}\right) \qquad \leq \varepsilon_c \tag{7a}$$

$$\varepsilon_z = \frac{\sigma_z}{E} \qquad \text{(compression)} \tag{7b}$$

$$\Delta D_a = \varepsilon_x \cdot D_a = \frac{\sigma_{th}}{E} \cdot \left(-\nu \cdot \frac{\sigma_z}{\sigma_{th}}\right) \cdot D_a \tag{7c}$$

$$\Delta D_{a-z} = \varepsilon_z \cdot D_a = \frac{\sigma_z}{E} \cdot D_a \qquad \text{(compression)} \qquad (7d)$$

The negative sign is used in Eq. (7) because the compressive stress $\sigma_z$ is negative. Consider the {100} crystal orientation so $\sigma_{th} = 26.4$ GPa, $E = 130$ GPa and $\nu = 0.279$ as in Table 1. If a uniaxial compressive stress $\sigma_z = 40$ GPa is applied onto SCS, the elastic strain $\varepsilon_x = 8.6\%$ is predicted by Eq. (7a), leading to the elastic deformation $\Delta D_a = 0.020$ (nm) from the atomic diameter $D_a$ of 0.235 (nm). For comparative purposes, assume the three energy bands of the Bohr atom in Fig. 1(a) have the same width. The estimated electron band width $\approx 0.235/6 = 0.039$ (nm). The lateral elastic deformation of a silicon atom in the x and y directions is already about a half of the band width. The compressive deformation in the z-direction is - 0.072 (nm), - 30% of the atomic diameter. The estimated lateral expansion in both x-y directions and compressive deformation in the z-direction would have considerably changed the space and stress environments of the electron orbitals inside the elastically deformed atoms. Reactions of the atoms against further compression have to come from the reactions of the electron orbitals through some rearrangements and formation of metastable pressure-resistant phases. However, it should be mentioned that the above linear analysis for uniaxial compression (around - 30% deformation) has been overestimated due to the non-linear behaviour towards the lower limit $D_0$ illustrated in Fig. 2. The non-linearity will show up under high pressures so that a much higher pressure is needed to generate the z-direction deformation indicated by Eq. (7d).

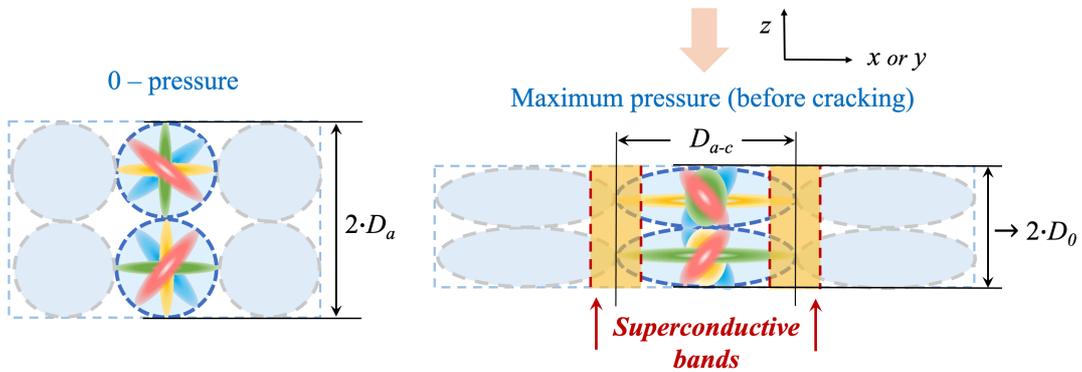

Figure 3. Electron orbitals in atoms are random before pressure is applied. Natural repulsive response of elastically deformed atoms suggests the electron orbitals are more aligned with the pressure around the nuclei (forming metastable pressure-resistant phases) to resist further compression towards the lower limit $D_0$. Consequently, the lateral elastically-expanded boundary region will contain fewer electrons, favourable for free electron movement and formation of superconductive bands.

Based on the tri-axial elastic strain analysis, a plausible mechanism for pressure-induced superconductivity can be suggested. The natural repelling response of atoms when being squeezed would mobilise more electrons towards the centre by aligning the previously random electron orbitals along the compressive direction, as illustrated in Fig. 3. Metastable pressure-resistant phases are generated [Lin et al. 1986, Zhang and Zarudi 2001, Kim et al. 2018]. Consequently, there would be fewer electrons in the lateral elastically expanded regions away from nuclei so that the electric conductivity were significantly enhanced around the atomic boundaries. It would be an ideal scenario that the applied pressure was so high that the majority of electrons was drawn to the nucleus along the z-direction to repel further compression. As a result, superconductive bands can be created at the boundary regions for free electron movement. So, the pressure-induced superconductivity is the result of extreme elastic deformation of atoms – rearrangement of electron orbitals around the nuclei – creation of superconductivity bands away from the nuclei.

The properties of nuclei cannot be changed easily under pressure, the transition from semi-conductor to conductor and finally to superconductor with increasing shear and pressure [Liu et al. 2020] can only come from rearrangements of electron clouds at the boundaries of atoms and inside the elastically deformed atoms. "The shear modes associated with the most severely deformation-weakened bonds", as commented in the study, must be linked to the elastic deformation around boundaries of diamond atoms. This may trigger a superconductive mechanism akin to what is explained in Fig. 3.

It should be emphasised, however, the elastic-strain based mechanism for pressure-induced superconductivity, suggested in Fig. 3, is highly speculative and only based on simple reasoning. Even nowadays it is still extremely difficult to observe electron orbitals within single atoms or around the boundaries [Chen et al. 2023]. The extremely fast and randomly orbiting electrons around nuclei could imply it may be impossible to clearly capture the electron orbital signatures. The challenge to observe electron movements highlights the significance of the work done by Pierre Agostini, Ferenc Krausz and Anne L'Huillier on experimental methods that generate attosecond pulses of light for the study of electron dynamics in matter. The real value of this unique mechanism for superconductivity and the concept of deformable atoms is that they may stimulate further research at the atomic scale. It is both interesting and challenging to picture the response of previously random electron orbitals in such inhospitable space and stress environments.

Close examinations of the AFM images [Chen et al. 2023] seem to suggest atoms are not 100% identical, so the reasoning of constantly oscillating atoms illustrated in Fig. 1(b) may not be too far away from the quantum world of atoms and electrons. For the time being, the elastic-strain based mechanism, illustrated in Fig. 3, at least provides a tentative mechanism for the pressure-induced superconductivity in silicon and similar solids. Possibly, there is some truth in this simple explanation, judging by Feynman's comments on atoms [1963] and the well-known relation between inter-atomic spacing and external stresses.

## 3. Elastic-Stain Enhanced Electric Conductivity and Superconductivity

It has been observed that the electric conductivity of silicon nanowires can be enhance under tensile and bending stresses [Shao et al. 2015]. Close to the critical elastic strain $\varepsilon_c$ under tension, electron orbitals will also be aligned along the tensile stress direction to avoid atom separation or fracture (if the same reasoning in Fig. 3 is followed). Although the widening of the electron energy band width along the tensile direction is favourable for electron movements, the superconductive bands created under compression as illustrated in Fig. 3 cannot be generated under tension or bending. It seems, up to now, reports on superconductivity are limited to extreme pressure conditions only.

Furthermore, the direct tensile or even bending strain can be easily limited by the atomic scale surface defects so it can be difficult to achieve a large fracture strain > 10%. For uniaxial compression experiments, extreme pressure can be applied around the centre of a plate sample away from the boundaries. In this way, the effects of surface defects are avoided. It should be emphasised the primary objective of this study is not to provide a comprehensive review on superconductivity and enhancement of electric conductivity due to elastic strains. Therefore, only few examples are selected and discussed below to show the present mechanics modelling is relevant to those phenomena.

The effects of tensile and bending strains on the electrical conductivities of silicon nanowires in the {110} direction have been measured at room temperature [Shao et al. 2015]. With around 1.5% tensile strain (before tensile fracture), the conductivity has been improved by 24%. Under the bending condition, the maximum strain on the tensile surface has reached to 5.8%, leading to 67% enhancement in the electrical conductivities. Using the direct tensile model, the increase in the atomic diameter $\Delta D_a$ = 1.5% x 0.235 = 0.0035 (nm) and $\Delta D_a$ = 5.8% x 0.235 = 0.014 (nm). The average electron band width is approximately around 0.039 (nm) ($\approx$ 0.235/6). So, the elastic strain of 5.8% under bending is not insignificant, and so is the 67% improvement in the electrical conductivity. Note as shown in Table 3 the critical elastic strain $\varepsilon_c$ can reach approximately 20% if there is no atomic-scale defect on the sample surface.

SCS has been tested under Vickers and Knoop indentations again at room temperature [Clarke et al. 1988]. Micro-cracking and very high pressures (close to 10 GPa) around the indenters have been generated. Under indentation load, the electrical conductivity of SCS has been improved by 2 orders of magnitude. The electrical conductivity has dropped almost to its initial level, if the indentation load is removed. Phase transformations to semimetal and metallic phases have been used to explain the significant change in the electrical conductivity. However, the lateral expansion in the x-y plane due to a high z-directional pressure as discussed in Section 2.2 can also be important.

The large elastic strain induced enhancement in the electrical conductivity at room temperature has also been observed in diamond nanoneedles under bending [Shi et al. 2020]. The maximum elastic strain around 10% has been generated on the tensile surface through nanoneedle bending. Again, the substantial elastic deformation of diamond atoms should not be excluded for the enhancement of electrical conductivity.

Pressure-induced superconductivity in silicon has been studied under pressures up to 43 GPa and temperature down to 2 K [Lin et al. 1986]. It is found that silicon is at a different phase for pressures from 35 to 43 GPa. So, the effects of high-pressure phases of silicon on superconductivity have been emphasized. As shown by the calculation using a uniaxial compressive stress $\sigma_z$ = 40 GPa in Section 2.2, the lateral elastic expansion in x-y direction is not insignificant and should definitely be included for explanations of the superconductivity phenomenon. To some degree, it is also possible that the superconductivity phenomenon in boron-doped diamond [Ekimov et al. 2004, Lee and Pickett 2004] is also linked to the elastic deformation around the doped atoms. Possibly, a mechanism akin to what is explained in Fig. 3 may be linked to the phenomenon.

It should also be mentioned that unconventional superconductivity without extreme pressure and elastic deformation of atoms has also been reported. One recent example is the "magic-angle" twisted bilayer graphene [Cao et al. 2018]. Since alteration of the electron orbitals in carbon atoms through elastic deformation is not involved, there must be other mechanisms that can promote free electron movements through those seemly normal carbon atoms. Hypothetically, one can assume two layers of small graphene samples can be twisted and pressed sufficiently close to each other in experiments so that the gap between them is measured in sub-nanometre, e.g., around the carbon atom diameter or closer. An effective superconductive band between the two graphene layers could be formed because of the random electron movements, which could act like a widened energy band for relatively free electron movement. If it were the case, the "magic angle" could be associated with the surface topography of graphene. In fact, there are publications on the two-dimensional fractional topological superconductor considering proximity effects [Qi et al. 2010, Vaezi 2013, Lu et al. 2020]. To some degree, they could be relevant to the magic-angle in graphene. One thing is clear the superconductivity phenomenon can only happen when the two layers of graphene are sufficiently close to each other.

## 4. Discussions

### 4.1 Atomic boundary and elastic deformation of atoms

By its definition, the interatomic spacing is the equilibrium distance between the nuclei of two atoms without experiencing external stress. The two-atom model adopted for silicon in this study has a strong covalent bond at the atomic boundary about the halfway between the two nuclei. A moderate tensile or compressive stress can only increase or decrease the interatomic spacing rather than breaking the covalent bond. Therefore, the atomic boundary formed by the covalent bond is still intact while the interatomic spacing is increased or decreased under the moderate stresses. If the above reasoning is correct, elastic deformation of the two atoms must have occurred with the changing interatomic distance. This has to be true if the space from the atomic boundary to the nucleus is considered as part of the silicon atom. Considering the three-dimensional space and random electron movements, atoms are not necessarily perfect spheres. The space-filled atomic model [Wikipedia, Space-filling model] shares some common features with the deformable atoms illustrated in Fig. 2.

Actually, the common relation between attractive and repulsive stresses and interatomic spacing has been known for many years. "Elastic deformation" of atoms has not been an issue since the interatomic spacing has been commonly interpreted as the interatomic "separation". Small "rigid atoms" (some distance apart) are connected by mysterious attractive and repulsive forces or waves so that there is plenty of room for the interatomic spacing to vary either in tension or compression. Following this above common interpretation of the interatomic spacing, the random electron orbitals inside an atom described by the Schrodinger quantum model cannot play any role in the atomic-scale attraction and repulsion. Clearly, this is not correct. The mysterious attractive and repulsive forces or waves must be part of the random electron orbitals and the atomic boundary formed by the covalent bond must exist around the mid distance between two nuclei.

The non-linear compressive behaviour due to intensified repulsion is also well recognized already without relying on the concepts of deformable atoms and lower limit $D_0$. However, if the atomic boundary is not clearly defined, it is difficult to bring the random electron orbitals and electron energy levels from the Schrodinger and Bohr models into the analysis. If the compressive deformation of atoms is accepted, formation of metastable pressure-resistant phases from electron rearrangements to resist further compression will come naturally. The electron structure variations inside atoms under extreme pressure should be linked to the pressure-induced superconductivity phenomenon.

The diameter of silicon atoms is only 0.235 nm. While it can be hard to imagine such tiny atoms shown in Fig. 1(a) are elastically deformable, it will be incomprehensible if they cannot be deformed under pressure. If the silicon atoms were not deformable, there would be no reduction in the interatomic spacing under extreme pressure. Also, there would be no increase in the interatomic spacing under tension before the complete separation of atoms. No elasticity for bonded atoms is clearly unrealistic and is not consistent with the well-known relation between attractive and repulsive stresses and interatomic spacing. Since the space inside an atom is either virtually empty or filled by random electron energy waves, atoms are not rigid and should be deformable in principle.

If the concept of deformable atoms is accepted, the door is open for new exploration. Any change in the three-dimensional space around a nucleus due to tensile, compressive and shear stresses will generate specific reactions from electrons accordingly so that new metastable variants (or elastically deformed atoms) with unforeseen characteristics and properties are created. This has to be the prime objective of Elastic Strain Engineering (ESE) at the atomic scale.

Of course, much simplified and bold assumptions have been made in this study to come to such a simple elastic-strain based explanation for pressure-induced superconductivity. Clearly, more studies will be needed to prove the relevance of those assumptions to the quantum world of atoms and electrons, and ultimately the press-induced superconductivity. Nevertheless, the current study at least has initiated a first attempt on modelling of elastic deformation of atoms, and provided simple formulas that can be easily tested by new experiments and compared with more focused atomic scale observations.

4.2     Griffith theory and four essential properties linked to the critical elastic strain $\varepsilon_c$

There are four major physical properties that are relevant to the critical elastic strain $\varepsilon_c$ and elastic deformation of atoms. They include the atomic diameter $D_a$, elastic modulus $E$, surface energy $\gamma_s$ and the theoretical strength $\sigma_{th}$, as can be seen from Fig. 2. The classic Griffith theory proposed over 100 years ago [1921] contains two of them, i.e., the elastic modulus $E$ and surface energy $\gamma_s$. By coincidence, the remaining two key parameters, the atomic diameter $D_a$ and theoretical strength $\sigma_{th}$, are contained in Eq. (3). Interestingly, Griffith theory and Eq. (3) do not share or possess a common physical property, but together they have covered all the necessary material parameters essential for prediction of the critical elastic strain $\varepsilon_c$, as illustrated in Fig. 2. The dominant characteristic length $L$ in Eq. (1) or the atomic diameter $D_a$ in Eq. (3) takes the elastic strain evaluation to the atomic scale.

The reliability of the simple and important relations specified by Eq. (4) and (5) depends on the accuracy of Eq. (3) because the classic Griffith theory has been extensively verified over 100 years and well accepted. To confirm the reliability of Eq. (3), fracture toughness predictions for three different materials (silicon, diamond and a superlattice with nano-scale ultra-thin layers) are selected and compared with experimental measurements from various sources. Such confirmations are necessary to bridge the two theoretical models established 100 years apart.

Table 4. Fracture toughness of silicon and diamond predicted by Eq. (3)

| Crystal planes | Silicon: $D_a$ = 0.235 (nm) | | Diamond: $D_a$ = 0.154 (nm) | |
|---|---|---|---|---|
| | $\sigma_{th}$ (GPa) | $K_{IC}$ (MPa√m) | $\sigma_{th}$ (GPa) | $K_{IC}$ (MPa√m) |
| {100} | 26.4 | 1.40 | 225 | 9.67 |
| {110} | 17.0 | 0.903 | 130 | 5.59 |
| {111} | 21.0 | 1.12 | 90 | 3.87 |

The fracture toughness $K_{IC}$ of silicon and diamond at different crystal planes predicted by Eq. (3) are listed in Table 4. A comprehensive review of silicon fracture [DelRio et al. 2015] noticed there were large scatters (around 0.6 to 2.5 MPa√m) in measured and predicted $K_{IC}$. After careful considerations, a "fortuitous" $K_{IC}$ range from 0.75 to 1.30 MPa√m was estimated for SCS, but the relevant crystal planes were not specified. The $K_{IC}$ predictions for SCS (from 0.90 to 1.40 MPa√m) in Table 4 are well covered by the estimated $K_{IC}$ range for SCS. The near effortless calculations using Eq. (3) based on two well-defined physical properties, $\sigma_{th}$ and $D_a$, at the atomic scale make the simple predictive method attractive.

The theoretical strength $\sigma_{th}$ of diamond in Table 4 is from Telling et al. [2000] and the atomic diameter $D_a$ is assumed to be the nearest neighbour distance [Sinclair and Lawn 1972, Field 2012]. $K_{IC}$ of diamond is then predicted using Eq. (3) and listed in Table 4. Drory et al. [1995] summarised diamond fracture toughness results from various sources (4 – 7 MPa√m without specifying crystal plane) and their own measurements from 4.9 – 6.2 MPa√m. Ferdous and Adnan [2017] reported $K_{IC}$ of diamond from the literature (4.7 – 14 MPa√m again without specifying crystal plane) and their own measurements around 8.85 MPa√m using various methods. The predictions of Eq. (3) listed in Table 4 are well within these two reported groups. It is also noticed that the fracture toughness $K_{IC}$ of diamond can be as low as 3.0 to 3.7 MPa√m in some collections of diamond material properties [AZO Materials – Diamond data sheet, MatWeb Material Property Data, websites in reference list]. The combined $K_{IC}$ scatter band for all the data from the above sources is as wide as 3 – 14 MPa√m and without a clear indication of the crystal orientation. The predicted fracture toughness $K_{IC}$ values, 3.87 – 9.67 MPa√m, in Table 4 for diamond with well-specified crystal planes provide valuable theoretical confirmations of the experimental measurements. Once again, the difficulty and uncertainty in $K_{IC}$ measurements for diamond makes Eq. (3) a valuable tool for independent assessments of the fracture toughness. Particularly, diamond samples are expansive and typically small so that fracture tests may be strongly influenced by the boundary effects if cracks are too close to the edges [Hu and Wittmann 2000, Zhang, Hu et al. 2018, Hu et al. 2022].

One useful application of different $K_{IC}$ values in Table 4 is that they can provide useful indications on the most likely cracking path in silicon and diamond. Nie et al. [2019] observed atomic scale defects on the surface of diamond nanoneedles, as shown in Fig. 4(a), marked by the 1 or 3 atom steps. They noticed that cracking of those diamond nanoneedles under bending invariably occurred along the {111} plane, as in Fig. 4(b). As can be seen from Table 4, the fracture toughness of 3.87 MPa√m along the {111} crystal plane is the lowest for diamond. Thus, it is logical to expect the preferred cracking path is always along the {111} facets, as shown in Fig. 4(b).

The fracture strain of those diamond nanoneedles shown in Fig. 4 [Nie et al. 2019] in the {100} direction varied from 6.5% to 13.4% and the flexural strength was as high as 125 GPa in the {100} direction. If the following data from Field [2012] are adopted for the {100} crystal direction, including the surface energy $\gamma_s$ = 9.2 Jm$^{-2}$, Poisson's ratio $\nu$ = 0.1 and elastic modulus $E$ = 1050 GPa and the atomic diameter $D_a$ = 0.154 nm (plus $\sigma_{th}$ = 225 GPa), the predicted critical elastic strain $\varepsilon_c$ of diamond from Eq. (5a) > 4.4%. The fracture toughness $K_{IC}$ estimated by the Griffith theory or Eq. (2) is 4.4 MPa√m, while $K_{IC}$ from Eq. (3) is 9.67 MPa√m in the {100} direction. If the correction factor $C_K$ > 1 is considered in Eq. (4a), the critical elastic strain $\varepsilon_c$ of diamond in the {100} crystal direction > 10%. Because diamond nanoneedles always contain atomic scale surface defects, most fracture strain measurements from bending tests are between 3.5 – 9% in Banerjee et al. [2018]. The fracture strain measurements from direct tensile tests are between 4.5 – 10% in {100}, {101} and {111} directions [Dang et al. 2021].

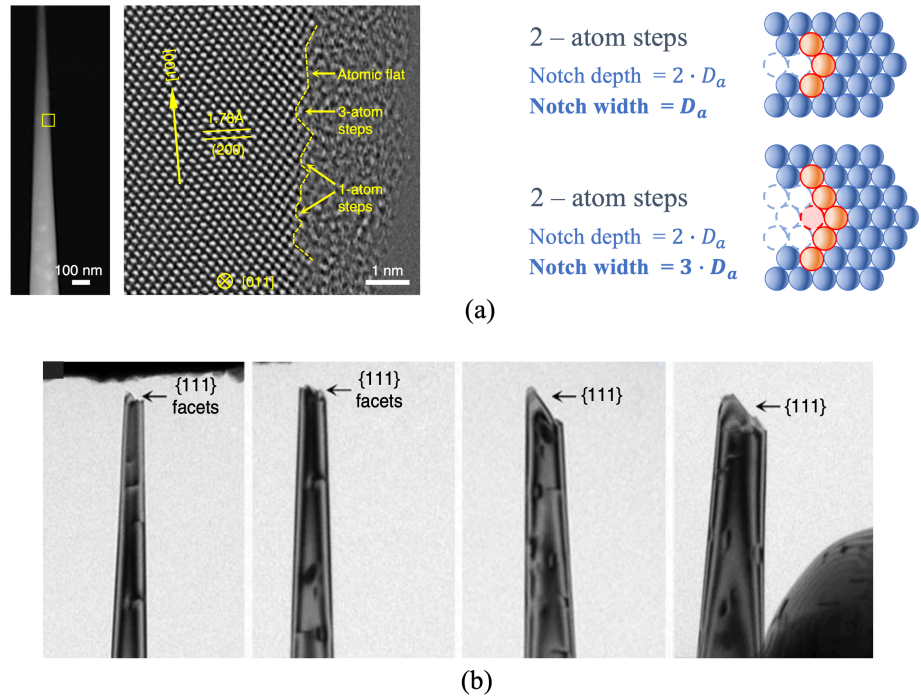

(a)

(b)

Figure 4. (a) A diamond nanoneedle is shown with an enlarged image from the circled surface section [Nie et al. 2019]. The atomic scale surface defects are denoted by the atom steps ($a_{a-s}$), that can be modelled by atomic notches with specified depth and width [Hu et al. 2022, Hu 2022]. (b) Under bending, fracture of diamond nanoneedles always occurred along the {111} crystal orientation. As can be seen from Table 4, $K_{IC}$ of 3.87 MPa√m for this crystal plane is the lowest.

It was speculated that the dominant characteristic length $L$ in Eq. (1) could even be a thin film thickness [Li et al. 2014]. Similarly, a combined nano-layer thickness can be used to replace the atomic diameter $D_a$ in Eq. (3) for fracture toughness predictions of nano-layered superlattices [Chen et al. 2023]. If the combined nano-layer $WN_x$/TiN (5 nm/5 nm) in the $WN_x$/TiN superlattice (or total thickness of around 10 nm) is used to replace $D_a$ and the average flexural strength from three measurements around 8.5 GPa is used to replace the theoretical strength $\sigma_{th}$, Eq. (3) predicts the fracture toughness $K_{IC}$ of the $WN_x$/TiN superlattice is around 3.0 MPa√m. Most fracture toughness measurements from the multilayered superlattices are between 2 and 4.5 MPa√m [Chen et al. 2023].

4.3 Elastic strain and deformation limited by atomic-scale defects

The critical elastic strain $\varepsilon_c$ of atoms and single crystals follows the basic relation $\sigma_{th} = E \cdot \varepsilon_c$. If pre-existing atomic-scale defects exist, the fracture strength $\sigma_f$ will be lower than the theoretical strength $\sigma_{th}$. As a result, the fracture strain $\varepsilon_f$ will be smaller than the critical elastic strain $\varepsilon_c$. It may come as a surprise to physicists and materials scientists that the effects of atomic-scale defects on the fracture strength and strain under tensile and bending conditions can be easily estimated using a simple mechanics model.

The influence of atomic scale surface defects, or the atom step $a_{a-s}$ in Fig. 4(a), on the fracture strength $\sigma_f$ of diamond can be estimated approximately by atomic notch models with notch depth and width linked to the atomic diameter $D_a$ [Hu et al. 2022, Hu 2022].

$$\sigma_f = \frac{\sigma_{th}}{\sqrt{1+\frac{a_{a-s}}{2 \cdot D_a + n \cdot D_a}}} \qquad n = 1, 2, 3, \ldots \tag{8a}$$

For nanocrystalline ceramics with a nano-scale surface crack $a$, the above relation can be rewritten using the intrinsic tensile strength $f_t$ (without influence of nano-/micro-defects) and average grain size $d_G$ [Zhang, Hu et al. 2018, Hu 2022].

$$\sigma_f = \frac{f_t}{\sqrt{1+\frac{a}{3 \cdot d_G}}} \tag{8b}$$

The notch width effect is not considered in Eq. (8b) but can be included if necessary [Hu 2022].

The two illustrative atomic notch models illustrated in Fig. 4(a) have the atomic notch depth or atom step $a_{a-s} = 2 \cdot D_a$ and the atomic notch width $N_w = n \cdot D_a$ with $n = 1$ and 3 respectively. It can be obtained from Eq. (8a) that the strength ratio $\sigma_f/\sigma_{th} = 0.77$ and 0.85 respectively, leading to around 15 to 23% strength reductions. These two cases show as long as atomic scale surface defects exist, even just one or a few atom steps, the measured fracture strength $\sigma_f$ will be lower than the theoretical strength $\sigma_{th}$ due to the stress concentrations created by the atom steps. Besides the above fracture strain and strength issues concerning the atom step surface topography, interactions of atom steps with a topological quantum state has also been investigated recently [Hossain et al. 2024], which is relevant to topological insulators and superconductors [Qi and Zhang 2011].

If atomic scale defects cannot be ruled out, a flexural test tends to yield a higher fracture strain than that from a direct tensile test due to the statistical reasons. While a uniform tensile test is always influenced by the maximum surface defect, the chance to find it around the maximum bending stress location is small. As expected, direct tensile tests of diamond produced smaller fracture strains of 5-9% [Dang et al. 2021] than the fracture strains, up to 13% [Nie et al. 2019], and around 10% [Banerjee et al. 2018] from flexural tests. The major issue for both tensile and flexural tests with atomic scale surface defects is to understand the interactions between the atomic surface defects and boundary conditions of test samples. In fact, the general issue on interactions between shallow surface defects and fracture strengths of large specimens (in comparison with surface defects) have been studied for many years by the "boundary effect model" [Hu and Wittmann 2000, Zhang, Hu et al. 2018, Hu et al. 2022] for various brittle solids with the characteristic microstructure $C_{ch}$, or $L$ in Eq. (1), from around 0.2 nm all the way up to around 100 mm.

4.4   Characteristic length $L$ and Hall-Petch relation

The dominant characteristic length $L$ in Eq. (1) plays a significant role in determination of the critical elastic strain $\varepsilon_c$ of a material, which is the focus of Elastic Strain Engineering (ESE). As mentioned earlier, it is noticed that the grain size in the Hall-Petch relation for metals [Hall 1951, Petch 1953] plays a similar role in determination of the yield strength. To show its relevance to the classic Hall-Petch relation, Eq. (3) is rewritten in a more general form using the characteristic microstructure $C_{ch}$, akin to $L$ in Eq. (1). The intrinsic tensile strength $f_t$ of a brittle polycrystalline solid, without the influence of defects, is used to as "the theoretical strength".

$$K_{IC} = 2 \cdot f_t \cdot \sqrt{3 \cdot C_{ch}} \tag{9a}$$

or

$$f_t = 0 + \frac{K_{IC}}{2\sqrt{3}} \cdot \frac{1}{\sqrt{C_{ch}}} \tag{9b}$$

The characteristic microstructure $C_{ch}$ can be the average grain size of a nanocrystalline ceramic [Hu 2022] or simply the atomic diameter $D_a$ of SCS or diamond as in this study.

The Hall-Petch relation [Hall 1951, Petch 1953] links the yield strength $\sigma_Y$ of a metal to its average grain size $d_G$ as follows.

$$\sigma_Y = \sigma_0 + k \cdot \frac{1}{\sqrt{d_G}} \qquad (20\ nm\ <\ d_G\ <\ 100\ \mu m) \qquad (10)$$

In this case, the average grain size $d_G$ is the dominant characteristic length $L$ or the characteristic microstructure $C_{ch}$ in E. (9). Therefore, the yield strength $\sigma_Y \propto L^{-0.5}$ with the exponent $\alpha = 0.5$. It is noted that the exponent $\alpha$ is between 0.5 and 1.0 [Li et al. 2014]. It seems $\alpha$ can indeed be either 0.5 or 1.0, depending on whether the critical stress or critical strain is considered.

The two fitting parameters, $\sigma_0$ and $k$, need to be determined experimentally through curve-fitting for a given metal. As illustrated in Fig. 5(a) [Yip 1998, Armstrong 2014], if the grain size $d_G$ is smaller than 20 nm, a reverse trend is observed ($\sigma_Y$ is decreased for nano-grained metals) so that the lower limit of the Hall-Petch relation is specified in Eq. (10) and Fig. 5(a).

Comparing Eq. (9b) and Eq. (10), it is clear that Eq. (9) in fact is the "Hall-Petch" relation for brittle solids [Hu and Wu 2024]. The new "Hall-Petch" relation is simpler ($\sigma_0 = 0$) and is well-defined as $K_{IC}$ is not a fitting parameter, but a material constant. It has been confirmed that Eq. (9) is valid for a wide range of brittle solids and composites with the characteristic microstructure $C_{ch}$ varying from the atomic scale around 0.2 nm (silicon and diamond) to close to 100 mm for large engineering structures such as dam concrete and brick walls [Hu et al. 2022]. As can be seen from the comparison of Fig. 5(a) and 5(b), the lower and upper limits of Eq. (9) are about 100 time smaller and 1000 time larger in comparison with those of the classic Hall-Petch relation of Eq. (10). Importantly, the new "Hall-Petch" relation for brittle solids, Eq. (3) or (9), has played a significant role in the derivation of the critical elastic strain $\varepsilon_c$ of atoms together with the Griffith theory.

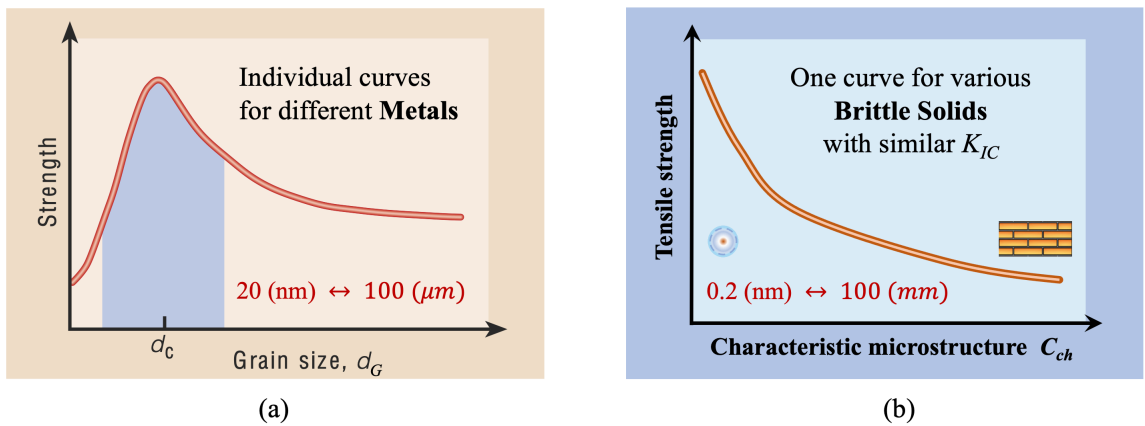

(a) (b)

Figure 5: (a) Hall-Petch relation for metals [Yip 1998], one curve for one metal only. The critical grain size $d_c$ for the transition is around 20 (nm). (b) Monotonic "Hall-Petch" relation for brittle solids with similar fracture toughness $K_{IC}$. One curve for all brittle materials as long as they have similar $K_{IC}$. The characteristic microstructure $C_{ch}$ can be as small as the atomic diameter $D_a$ for single crystal silicon and diamond ($\approx 0.2$ nm) or as big as height of bricks (over 60 mm) [Wang et al. 2020].

## 5. Concluding Remarks

Change in the interatomic spacing of a two-atom system under tension and compression has been modelled by the elastic deformation of atoms. The hypothesis, that atoms can be deformed elastically under stress, has been formulated by two mechanics models proposed a century apart. The predicted critical elastic strain $\varepsilon_c$ around 10% or larger for both silicon and diamond atoms under tension has been confirmed by sophisticated tensile and flexural experiments using carefully fabricated single crystal nanoneedles. The hypothesis of deformable atoms has led to a unique elastic-strain based mechanism for superconductivity originated from the lateral elastic expansion and alignment of previously random electron orbitals along the pressure direction. This in fact is consistent with the natural repulsive response of atoms against further compression. Theoretically, superconductive bands can be created away from the nuclei in the lateral elastically expanded boundary regions, where there is more space for resistance-free electron movement.

However, it should be emphasised that the elastic-strain based mechanism for pressure-induced superconductivity is highly speculative and much simplified. Many factors including phase transformation and non-linear compressive deformation can also influence the complex phenomenon of superconductivity. This elastic-strain based mechanism together with the hypothesis of deformable atoms should therefore be treated with caution. The real value of this unique mechanism for superconductivity and the concept of deformable atoms lies in the possibility they may stimulate further research in the area of Elastic Strain Engineering (ESE) at the atomic scale. It is both interesting and challenging to picture the response of previously random electron orbitals in such inhospitable space and stress environments.

This study has verified Eq. (1), important to ESE, is valid at the atomic scale. The "Hall-Petch" relation for brittle solids Eq. (3) or (9), used to confirm Eq. (1), is applicable from the atomic scale all the way up to engineering structures. The fact that such a simple formula can cover such a wide range of materials warrants further investigation. The unexpected prediction of the critical elastic strain $\varepsilon_c$ of atoms from Eq. (3) together with the Griffith theory proposed a century ago is a good example.

Finally, it can be imagined that in the quantum world of atoms and electrons, nuclei resonate with fast orbiting electrons so that atoms in matter are constantly oscillating around their equilibrium positions in the form of elastic deformation. Living organisms on earth are adapted to their environment in order to thrive. Atom's natural repulsive reaction to extreme pressure, through mobilizing electrons to form pressure resistant metastable phases, is no different. The willingness of atoms to deform elastically in a harmonic oscillation with quantum-state electrons, plus their ability to resist external stresses through rearranging previously random electron orbitals and the instinct to change their bonds with other atoms whenever deemed necessary because of the changing environment, is probably the reason why our world is so diverse and colourful.

**Acknowledgements**

This study was partially supported by the Australian Research Council Discovery Grant DP170104307 (June 2017 – December 2023), Predicting strength of porous materials: A microstructure-based approach.